\documentclass[conference]{IEEEtran}

\IEEEoverridecommandlockouts
\usepackage{soul}
\usepackage{atbegshi}% http://ctan.org/pkg/atbegshi
\AtBeginDocument{\AtBeginShipoutNext{\AtBeginShipoutDiscard}}
\usepackage{eso-pic}

\usepackage[inline]{enumitem}
\usepackage{amsmath}
\usepackage{bm}
\usepackage{amssymb}
\usepackage{soul}
\usepackage{graphicx}
\usepackage{comment}
\usepackage{color}
\usepackage{subcaption}
\usepackage{multirow}
\usepackage{tabularx,booktabs}

\usepackage{algorithm}%,algorithmic}
\usepackage[noend]{algpseudocode}

\usepackage{outlines}

\usepackage{caption}
\usepackage{subcaption}

\algnewcommand{\LineComment}[1]{\State \(\triangleright\) #1}

\usepackage{array}
\newcolumntype{L}[1]{>{\raggedright\let\newline\\\arraybackslash\hspace{0pt}}m{#1}}
\newcolumntype{C}[1]{>{\centering\let\newline\\\arraybackslash\hspace{0pt}}m{#1}}
\newcolumntype{R}[1]{>{\raggedleft\let\newline\\\arraybackslash\hspace{0pt}}m{#1}}

\usepackage{mathtools}

\DeclarePairedDelimiterX\set[1]\lbrace\rbrace{#1}

\definecolor{dkgreen}{rgb}{0,0.6,0}
\definecolor{gray}{rgb}{0.5,0.5,0.5}
\definecolor{mauve}{rgb}{0.58,0,0.82}

\ifCLASSINFOpdf
\else
\fi
\begin{document}

\def\LWn{\dimexpr.44\linewidth}
\def\LW{\dimexpr.485\linewidth}%-.5em}
\def\LWW{\dimexpr.005\linewidth}%-.5em}


%\title{JupiterTP - A System for Optimizing the Throughput of Dispersed Computing}

\title{Design and Experimental Evaluation of Algorithms for Optimizing the Throughput of Dispersed Computing}

\thanks{}

\author{\IEEEauthorblockN{Xiangchen Zhao}
\IEEEauthorblockA{%Ming Hsieh Department of Electrical Engineering\\
University of Southern California\\
% Los Angeles, California 90089\\
zhao115@usc.edu}
\and
\IEEEauthorblockN{Diyi Hu}
\IEEEauthorblockA{%Ming Hsieh Department of Electrical Engineering\\
University of Southern California\\
% Los Angeles, California 90089\\
diyihu@usc.edu}
\and
\IEEEauthorblockN{Bhaskar Krishnamachari}
\IEEEauthorblockA{%Ming Hsieh Department of Electrical Engineering\\
University of Southern California\\
% Los Angeles, California 90089\\
bkrishna@usc.edu}}

% =====================================================================


%Telephone: (800) 555--1212, Fax: (888) 555--1212}


\maketitle

%\AddToShipoutPicture*{\small\sffamily\raisebox{1.8cm}{\hspace{1.8cm}978-1-5386-3797-5/17/\$31.00 \copyright2017 IEEE}} 


% \fi
\IEEEpeerreviewmaketitle

\begin{abstract}
% overall context
With growing deployment of Internet of Things (IoT) and machine learning (ML) applications that need to leverage computation on networked edge and cloud resources, it is important to develop algorithms and tools to place these distributed computations to optimize their performance. We address the problem of optimally placing computations described as directed acyclic graphs (DAGs) over a given network of computers, to maximize the steady-state throughput for pipelined inputs. Traditionally, such optimization has focused on a different metric, minimizing single-shot makespan, and a well-known algorithm is the Heterogeneous Earliest Finish Time (HEFT) algorithm. Keeping in mind the objective of maximizing throughput which is more suitable for many real-time, cloud and IoT applications, we present a different scheduling algorithm that we refer to as Throughput HEFT (TPHEFT). Further, we present two throughput-oriented enhancements which can be applied to any baseline schedule, that we refer to as ``node splitting" (SPLIT) and ``task duplication" (DUP).  In order to implement and evaluate these algorithms, we built new subsystems and plugins for an open-source dispersed computing framework called Jupiter. Experiments with varying DAG structures indicate that: 1) TPHEFT can significantly improve throughput performance compared to HEFT (up to 2.3 times in our experiments), with greater gains when there is less degree of parallelism in the DAG, 2) Node splitting can potentially improve performance over a baseline schedule, with greater gains when the baseline schedule has an imbalanced allocation of computation or inter-task communication, and 3) Task duplication generally gives improvements only when running upon a baseline that places communication over slow links. To our knowledge, this is the first study to present a systematic experimental implementation and exploration of throughput-enhancing techniques for dispersed computing on real testbeds. 

% overall context
% specific problem being addressed
% new idea / solutions proposed
% what is the hypothesis that is evaluated 
% how the evaluation is done
% what the results are
    
\end{abstract}


\section{Introduction}
Dispersed computing, where complex applications that process data from multiple sources are distributed across networked computers, is a promising paradigm for real-world applications~\cite{19middleware, dcomp_intro, coded_dispersed, social_dispersed}. Like real-time video processing and cloud computing, there’s often a massive amount of data continuously and steadily fed into the system from multiple sources. In today’s big data applications, the amount of compute, storage and network resources required to finish a job in a reasonable amount of time far exceeds those of a single computer and therefore a cluster of compute nodes is needed. However, when computation is not centralized on one machine, various challenges emerge, such as scheduling~\cite{scheduling_cloud}, coordination, fault tolerance~\cite{fault_tolerance} etc. Besides, machines are usually commodity hardware, and have heterogeneous performance when processing different types of jobs~\cite{mobile_benchmark, cpu_benchmark}. Inter-machine communication is also heterogeneous due to non-linear relationship between transfer time and file size~\cite{Jupiter_Quynh}. 

Distributed applications are often described by directed acyclic graphs (DAG). Each application DAG consists of multiple tasks with precedence requirements. Each input instance needs to go through the exact same process flow of the DAG, and generates an output. Traditionally, in grid computing, the objective is often minimizing the single-shot makespan, such as the famous HEFT schedule~\cite{heft}. However, for dispersed computing involving processing of streaming data sources such as in IoT applications, the computation can be pipelined and thus a more useful parameter is throughput, defined as the number of instances processed per unit time. In this work, therefore, instead of minimizing makespan for a single data input, we consider maximizing the steady-state throughput for pipelined execution.

Google invented Kubernetes~\cite{borg} (k8s) to manage container orchestration on large scale clusters. However, k8s doesn't have a scheduler that implements a globally optimized algorithm; nor does it have a comprehensive orchestration system to process pipelined DAG applications. 

Keeping in mind these challenges, we propose three throughput optimized scheduling algorithms, namely ``Throughput HEFT" (TPHEFT), ``Node Splitting" (SPLIT) and ``Task Duplication" (DUP), and implement a real-world system (JupiterTP) based on k8s and our previous Jupiter system~\cite{Jupiter_Quynh} to perform end-to-end pipelined computation on a distributed cluster. Most prior work on throughput has focused on algorithms and simulations. To our knowledge, this is the first work to not only propose algorithms, but also implement a full-stack distributed system for throughput optimization, which can handle arbitrary DAGs, and can be applied over any network of computers --- public or private cloud, IoT systems, edge networks etc. The system itself is much more than just schedulers and scheduling algorithms, the profilers, execution engines and orchestration engines together make the above possible. Our system also provides unified interfaces to include other scheduling algorithms of similar settings, so that they can go beyond simulations. What's more, we systematically studied the performance of the whole system using real clusters, and also some real world applications, which is much more convincing than just simulated results. Our system has been made publicly available as open-source software.

\section{Related work}

The DAG and heterogeneous cluster scheduling problem for makespan minimization is NP-complete~\cite{npcomplete}. There have been many heuristics proposed, which can be roughly categorized into three classes: list scheduling~\cite{heft}, cluster scheduling~\cite{cluster_tpds, cluster2} and duplication scheduling~\cite{diyi, cluster_icpp}. 
%A list scheduling heuristic ~\cite{heft} maintains a list of all tasks sorted by the defined “priority”, and has the "task prioritizing phase" and the "processor selection phase". In clustering Heuristics~\cite{cluster_tpds, cluster2}, each step of iteration refines the previous clustering by merging some clusters. Through several iterations, the final number of clusters is equal to the number of processors. In Task Duplication Heuristics~\cite{diyi, cluster_icpp}, some tasks may be processed multiple times to reduce the transfer time. This is done by putting the parent task closer to its child, and is usually targeted for clusters with abundant resources. %
%Note that in this paper, our duplication is different from the usual makespan duplication, and details will be introduced in the algorithm part.

Though less studied, there has also been prior work on improving throughput.~\cite{ipdps09} proposed greedy and linear algorithms, and conducted simulation on SimGrid framework; ~\cite{odessa} proposed real systems, but for only 2 devices - mobile and cloud; there are also other real systems designed for special DAGs, such as linear DAGs in~\cite{linearDAG}, and the map-reduce framework~\cite{mapreduce}. Further work on throughput maximization such as~\cite{theory}~\cite{simu}~\cite{dup_tp_stream}~\cite{lat_tp} are theory or simulation based. We considered implementing their algorithms as baselines, but some assumptions in those papers are idealized don't hold in the real system, thus cannot be implemented easily and uniformly. Our work is different from previous ones, in a sense that we have novel algorithms and end-to-end full-stack systems for general DAGs.

Some pipelined processing systems (such as k8s itself) use dynamic scheduling, where the scheduler, based on the current status of the system, decides where to put the next task. This is best applied when we don’t have much knowledge (such as future workload), or the system has a high churn rate or unexpected behaviors. However, we're aimed at optimizing expected overall system performance, not for the current task. In our model, tasks are templates scheduled on nodes, they start their routines when an input instance arrives; and cluster nodes and links are working steadily with small variations. Given such knowledge, we can do a one-shot, globally optimized scheduling, which can be reapplied when the DAG or cluster status changes. Once we decide where to run a task, we run all input instances of this task on the same node until the next round of scheduling.

\section{System model and key concepts}
In this paper, we use the terms `node', `processor' and `proc' interchangeably, since we target single-core nodes. Tasks can’t run in parallel on single-core nodes, same for file transfer on links. We optimize for single-core nodes in theory, but though not as precise, our algorithms can also improve throughput for multi-core nodes.

\subsection{Application DAG and compute cluster}
In the general form of applications, a DAG consists of a set of tasks, represented by vertexes; and inter-task file transfers, represented by directed weighted edges.

A cluster is a collection of resources, and our main interest is computation and communication resources. Among these two, the former is represented by a set of nodes, consisting of commodity hardware, which can be highly heterogeneous, with nodes perform differently under different types of tasks. Some previous work used a simple model, where

%\begin{equation}
%    \textrm{execTime[task][node]} = \frac{\textrm{load[task]}}{ %\textrm{compFactor[node]}}
%\end{equation}

\begin{equation}
    execTime[task][node] = load[task] / comp[node]
\end{equation}

\emph{i.e.}, assigning a single compute factor to each node. However, as a counter-example, some nodes have powerful CPU while others perform well for I/O. If a task consumes a lot of CPU, it will perform better on the first type of nodes; if another task is I/O heavy, it'll perform better on the second type. Thus we can't simply assign a compute factor to each node. Instead we use a data-driven approach to profile the expected execution time of each task on each node, and describe execution as a matrix. 

On the other hand, communication is represented by a set of links connecting nodes. All nodes in the cluster are reachable to each other. In a regular commercial cluster, we don’t know or have control over the link topology or the routing tables, thus we use an abstraction of links, namely “virtual links”, and assume there is a pair of links between each pair of nodes. In reality, several virtual links can be overlapped as one actual (physical) link. We showed in later experiments how accurate our model is, and how to adjust parameters when the model is inaccurate. Similar to computation, most prior work states 
\begin{equation}
    time[file][link] = size[file] / bandwidth[link]
\end{equation}
In our work, inspired by the findings in~\cite{Jupiter_Quynh}, instead of assigning a fixed bandwidth to each link, we used a quadratic model
\begin{equation}
    time[file][link] = a \cdot (size[file])^2 + b \cdot (size[file]) + c
\end{equation}
where a, b, c are parameters to describe a link, calculated by quadratic regression of existing data of file sizes and transfer times.

\subsection{Example process to reach steady-state}
Single-shot makespan and steady-state throughput are different concepts and don't have a direct relationship. Steady-state means the system is working periodically, repeating the same process for each period. Thus minimizing makespan is not the same as maximizing throughput. In each schedule's steady-state, there's always a bottleneck resource which defines the overall system throughput. So we need to find ways to reduce the current bottleneck.

During an actual process, there’s usually a start-up stage, where the system is partly saturated, and after some time, when the first input instance reaches output, the system will reach a steady-state.

Fig~1 is a simple example to illustrate how a system can reach steady-state, and how to identify the bottleneck. Imagine a linear cluster that consists of three worker nodes connected by two links; and a DAG that consists of four tasks with a diamond shape. Given a schedule, T0 (task0) is on on node1; T1 and T2 on node2; T3 on node3. Assume the four tasks take 3, 2, 2, 5 seconds to complete (on their scheduled nodes respectively). Also assume the file transfer between T0 and T1, T0 and T2, T1 and T3, T2 and T3 takes 2, 1, 3, 2 seconds respectively (on their scheduled links). Neither node nor link can process more than one job at the same time. 

A detailed description of what each stage looks like is provided in table~I. We use “Tx-Ny” to denote input instance y for task(x), use “Fmn-Ny” to denote the file transfer from task(m) to task(n) for input instance y. We assume continued input instances N1, N2, N3... are generated fast enough. In table~I, during the first 15 seconds, parts of the system are idle with no fixed routine. After 20 seconds, when the first output (which belongs to input1) was generated, the system reaches steady-state, and generates an output every 5 seconds. The rate at which the outputs are generated (throughput) depends only on the system bottleneck, which is the critical resource working non-stop. Other resources can take a break in each time period. In this example, both node3 (handles one task in each period) and link23 (handles two file transfers in each period) are the system bottlenecks.

\begin{figure}[htbp]
\centering
\includegraphics[scale=0.35]{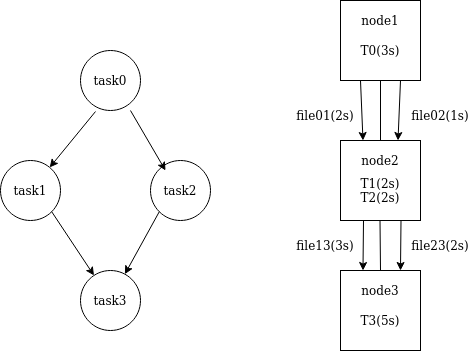}
\caption{Example of a simple DAG and schedule}
\end{figure}

% \usepackage{graphicx}
\begin{table}[]
\centering
\caption{Stages to reach steady-state}
\resizebox{0.47\textwidth}{!}{%
\begin{tabular}{|l|l|l|l|l|l|l|}
\hline
 &
  \begin{tabular}[c]{@{}l@{}}\\ node1\\\\ \end{tabular} &
  link12 &
  node2 &
  link23 &
  node3 &
  OUT \\ \hline
0-3s &
  \begin{tabular}[c]{@{}l@{}}\\ T0-N1\\\\ \end{tabular} &
   &
   &
   &
   &
   \\ \hline
3-6s &
  \begin{tabular}[c]{@{}l@{}}\\ T0-N2\\\\ \end{tabular} &
  \begin{tabular}[c]{@{}l@{}}F01-N1(2s)\\ F02-N1(1s)\end{tabular} &
   &
   &
   &
   \\ \hline
6-10s &
  \begin{tabular}[c]{@{}l@{}}T0-N3(3s)\\ idle(1s)\end{tabular} &
  \begin{tabular}[c]{@{}l@{}}F01-N2(2s)\\ F02-N2(1s)\\ idle(1s)\end{tabular} &
  \begin{tabular}[c]{@{}l@{}}T1-N1(2s)\\ T2-N1(2s)\end{tabular} &
   &
   &
   \\ \hline
10-15s &
  \begin{tabular}[c]{@{}l@{}}T0-N4(3s)\\ idle(2s)\end{tabular} &
  \begin{tabular}[c]{@{}l@{}}F01-N3(2s)\\ F02-N3(1s)\\ idle(2s)\end{tabular} &
  \begin{tabular}[c]{@{}l@{}}T1-N2(2s)\\ T2-N2(2s) \\ idle(1s) \end{tabular} &
  \begin{tabular}[c]{@{}l@{}}F13-N1(3s)\\ F23-N1(2s)\end{tabular} &
   &
   \\ \hline
15-20s &
  \begin{tabular}[c]{@{}l@{}}T0-N5(3s)\\ idle(2s)\end{tabular} &
  \begin{tabular}[c]{@{}l@{}}F01-N4(2s)\\ F02-N4(1s)\\ idle(2s)\end{tabular} &
  \begin{tabular}[c]{@{}l@{}}T1-N3(2s)\\ T2-N3(2s)\\ idle(1s)\end{tabular} &
  \begin{tabular}[c]{@{}l@{}}F13-N2(3s)\\ F23-N2(2s)\end{tabular} &
  T3-N1(5s) &
  N1 \\ \hline
20-25s &
  \begin{tabular}[c]{@{}l@{}}T0-N6(3s)\\ idle(2s)\end{tabular} &
  \begin{tabular}[c]{@{}l@{}}F01-N5(2s)\\ F02-N5(1s)\\ idle(2s)\end{tabular} &
  \begin{tabular}[c]{@{}l@{}}T1-N4(2s)\\ T2-N4(2s)\\ idle(1s)\end{tabular} &
  \begin{tabular}[c]{@{}l@{}}F13-N3(3s)\\ F23-N3(2s)\end{tabular} &
  T3-N2(5s) &
  N2 \\ \hline
25-30s &
  \begin{tabular}[c]{@{}l@{}}T0-N7(3s)\\ idle(2s)\end{tabular} &
  \begin{tabular}[c]{@{}l@{}}F01-N6(2s)\\ F02-N6(1s)\\ idle(2s)\end{tabular} &
  \begin{tabular}[c]{@{}l@{}}T1-N5(2s)\\ T2-N5(2s)\\ idle(1s)\end{tabular} &
  \begin{tabular}[c]{@{}l@{}}F13-N4(3s)\\ F23-N4(2s)\end{tabular} &
  T3-N3(5s) &
  N3 \\ \hline
\end{tabular}%
}
\end{table}

\subsection{How to identify bottlenecks}

Based on the above analysis, we define a term called ``schedule time" of a resource, which is the actual working time of a resource in one period, excluding idle times. The maximum schedule time of a system is the maximum value of the schedule times of all resources. For the bottleneck resource, its schedule time equals actual working time. Thus finding the system bottleneck is finding the resource with largest schedule time. Note that in general, we only consider nodes and links. There are other resources and possible bottlenecks, such as multi-socket overhead during file transfer startup, we will summarize them in the experiment part and give an overview of how much they cost. But based on our experiment results, only considering these two resources is a useful model under most circumstances.

Assume we have a schedule where node P has scheduled n tasks, which takes time t1, t2, ..., tn respectively, thus the schedule time of node P is $\sum_{i=1}^{i=n}(t_i)$. Assume file transfers on link L take time f1, f2, ..., fm respectively, then the schedule time of this link is $\sum_{i=1}^{i=m}(f_i)$. The relationship between achievable steady-state throughput and schedule times of nodes and links is as follows
\begin{equation}
    TP_{sys} = 1 / \max ( \max (T_{node}), \max (T_{link}))
\end{equation}

%\begin{equation}
%    TP_{sys} = \frac{1}{ \max ( \max (T_{node}), \max (T_{link}))}
%\end{equation}

\section{Algorithms}

All of our algorithms are designed to maximize the steady-state throughput, under an environment with abundant resources. We start from HEFT, analyze the bottlenecks, then propose TPHEFT, which is an independent algorithm that aims to minimize bottlenecks; then propose SPLIT and DUP, which can be applied to any existing algorithms to improve their throughput. 

\subsection{Throughput HEFT (Algorithm Block 1)}

In single-shot case, one key trick to minimize makespan is to reduce inter-task file transfer times. It is considered that on the same node, inter-task file transfer time is always zero, given that inter-process communications are much faster than network transfer. So HEFT tends to schedule tasks close to each other if they have dependencies. As an extreme case, for a singly linked-list DAG, HEFT will schedule all tasks on the same node. The only way HEFT can schedule tasks on different nodes is if some tasks can run in parallel. For a pipelined case, parent and child tasks can be processed at the same time using different input instances. In this case, HEFT clearly creates idle resources thus decrease throughput.

Thus we propose TPHEFT, whose overall goal is to minimize the maximum system schedule time across all resources, so that the bottleneck is minimized. There has been prior work to improve HEFT to boost throughput~\cite{ipdps09}. Our algorithm is different in terms of scheduling order, and also in each iteration we only minimize the schedule time over effected resources, which is more precise and can have better results. Similar to HEFT, we define task upward rank recursively as
\begin{equation} 
    rank(n) = comp(n) + \max{_{k\in{n child}}}(comm(nk) + rank(k)) 
\end{equation}
Where comp(n) is average computation cost of task(n) over all nodes, comm(nk) is average communication cost of file transfer from task(n) to task(k) over all links. Upward rank is a more precise topological ranking in weighted graphs. Scheduling in this order can also make sure that all parent tasks are already scheduled before scheduling a child task. Note that upward rank does not guarantee time dependency because it's not the precise value of task time on the scheduled node, so HEFT has another step to make sure the earliest start time of a child is later than the latest finish times of all parents. 

In TPHEFT, we drop the time dependency concern, and only use upward rank, simply because in pipelined execution, a child task can be executed before the parent task is finished, they're just at different time stages. For example, the child may be executing task(c)-input1 while the parent is executing task(p)-input2. In our implementation, we use run time queues and file system monitoring process to handle this.

Following upward ranks, we tentatively schedule the current task on a node. This will cause one incurred node usage, and several incurred link usages. We add these new costs to these resources' original costs, and calculate the resulted max schedule time over effected resources. This value means, if current task is scheduled on this node, the max schedule time of effected resources. Note that this is not necessarily the current system max schedule time. We pick such a node to minimize this value, and schedule current task on this node.

\begin{algorithm}
\caption{Throughput HEFT}
\begin{algorithmic}[1]
\renewcommand{\algorithmicrequire}{\textbf{Input:}}
\renewcommand{\algorithmicensure}{\textbf{Output:}}
\Require \\
Application DAG, sorted tasks \\
Cluster (undirected weighted graph) \\
Network profile (list of (a, b, c)) \\
Execution matrix, exec[i][j] is time of task i on node j 
\Ensure Mapping from tasks to nodes, one task to one node \\

\State $resTime\gets \{\}$ \Comment{map from resources to their schedule times, update when a task is scheduled}
\State assign $entryTask$ first, update resources
\ForAll{$task\in\mathit{tasks}$}
    \State {$tmp \gets \{\}$} \Comment{map from a node to resulted max schedule time over *effected* resources if $task$ is scheduled on this node}
    \State $pTasks, pNodes \gets$ parent tasks, nodes of $task$
    \ForAll{$proc\in\mathit nodes$}
        \State $nodeTime \gets$ exec[$task$][$proc$] + $resTime$[$proc$]
        \State $linkTime \gets$ copy existing link costs in $resTime$
        \ForAll{$pTask \in\mathit pTasks$}
            \State $link$ = $pNode$-$proc$
            \State $lTime$ = time of file $pTask$-$task$ on $link$
            \State $linkTime$[$link$] += $lTime$
        \EndFor
        \State $maxLinkTime$ = max $linkTime.values()$
        \State $maxTime$ = max($nodeTime$, $maxLinkTime$)
        \State $tmp$.add($proc$, $maxTime$)
    \EndFor
    \State assign $task$ to key (proc) in $tmp$ with smallest value
    \State update effected resource usages
\EndFor
\end{algorithmic} 
\end{algorithm}

\subsection{Node Splitting (Algorithm Block 2)}
\label{sec: split}

We propose SPLIT, an algorithm that aims to improve the throughput of an existing schedule by making portion-based replicas of all tasks on a node. Consider HEFT, TPHEFT, or any other schedule, the critical resources can be significantly more time consuming than others, especially in HEFT where tasks are more centralized. In TPHEFT, although we already tried to balance them, we're bounded by the condition ``one task can only be scheduled on one node". If we just happen to have a task which is very time consuming, there's nothing TPHEFT can do. Thus we propose SPLIT, where we reduce the bottleneck on nodes or links by splitting a node (meaning all tasks scheduled on this node) into several replicas, each with an assigned portion of original workload.

A runtime example of the algorithm:\\
Input: \{`task0': `node1', `task1': `node5'\} \\
Output:
\{`task0': [`node1':
           0.15,
           `node2':
           0.36,
           `node3':
           0.49],
 `task1': [`node5': 0.32, `node6': 0.68]\}

There are several challenges arising from this idea, in terms of algorithm, system implementation and experiments. Below are some key remarks.

\begin{itemize}
\item We can't require the application programmer to write their tasks in a multi-thread way, and processing one input instance of a task on multiple nodes for DAGs is very hard. In SPLIT, we still process each input instance on one node, but will dynamically distribute different instances of the same task to different nodes based on their portions, using both random number and hashcode-bucket methods.
\item If bottleneck is a link, we can do SPLIT on source or destination node. This will reduce the inter-task file transfer load between the two nodes, thus reducing the link bottleneck. So we only talk about node splitting.
\item When deciding where to split the bottleneck resource to, our number one priority is to NOT result in a bigger bottleneck, because when we incur extra load somewhere else, the new resource could become the new bottleneck. Our number two priority is to reduce the load on current node as much as possible.
\item When we split nodes, should we split it into two, three, or four? As mentioned before, improving anywhere other than the bottleneck is useless, and if at some point of splitting, this resource is not the bottleneck anymore, then further splitting it is wasteful.
\item SPLIT can always result in better throughput if handled properly, because we can control the portion of tasks to ship. Even shipping away a tiny portion will reduce the bottleneck.
\end{itemize}

We summarize the algorithm as two phases, the node and portion selection phase, and the do split phase. In case of link bottleneck, just split source or destination node. In phase one, first find the bottleneck node (btnk). Then iterate over all candidate nodes (We consider candidate nodes as idle nodes, simply because we have abundant resources. But this can be easily extended to any node with a small usage), for each candidate node, tentatively split 100\% btnk workload onto this node. Like in TPHEFT, this will cause one extra node usage and several extra link usages. Get the max schedule time over effected resources, mark it as the max schedule time if btnk is scheduled on this candidate. After checking all candidates, find the one with smallest max schedule time as the objective node. After deciding on this candidate, we choose a portion of btnk to ship to the candidate, such that the schedule time of btnk node and effected resources become the same. Note that the portion means the workload percentage of each task relative to btnk, not original tasks. Fig~2 gives an example that btnk has already been split and carries different portions of different tasks. Splitting 40\% of the node to a new node will result in a schedule in the fig. This way, we split each node into two for every iteration, but can keep splitting if the new nodes are still bottlenecks, which is more agile and simple than calculating ``splitting btnk into how many candidate nodes and what portions". Note that task portions and node portions are not much related, because a task is on several nodes, a node has several tasks. We only keep track of the portion (relative to original task) of each task on each node, as in the previous SPLIT output example.

\begin{figure}[htbp]
\centering
\includegraphics[scale=0.4]{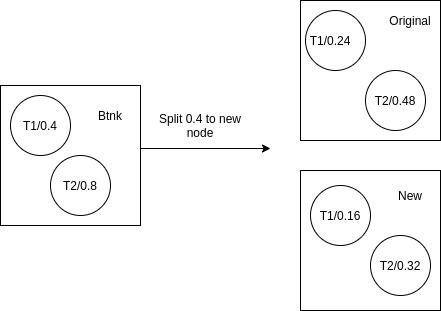}
\caption{Example of portion after split}
\end{figure}

In phase two, split btnk node into two nodes and update resource usages based on portion. Fig~3 describes an example case for SPLIT. Effected resources include: btnk node, candidate node, links between \{all parent and child nodes of btnk node\} and btnk node, links between \{all parent and child nodes of btnk node\} and candidate node. These resource usages need to be updated based on node splitting portion. Here parent nodes of nodeB are defined as nodes where any scheduled task is a parent of any scheduled task on nodeB. Cluster nodes are undirected graph and have no parent-child relationship originally, but after scheduling tasks on them, such relationships appear. Obviously, unlike in tasks, the patent-child relationships of nodes is not a DAG, a node can be a parent of itself, or nodes can form a cycle of parent-child relationships, or a node can have no parent or child nodes. These corner cases need to be carefully managed.

 Although SPLIT generated more tasks and parent-child relations, we do NOT consider this as a change of the original graph, instead we consider the extra tasks as replicas, such as taskA-1, taskA-2, etc, which is different from task duplication in the next section.

\begin{figure}[t]
    \centering
    \includegraphics[scale=0.32]{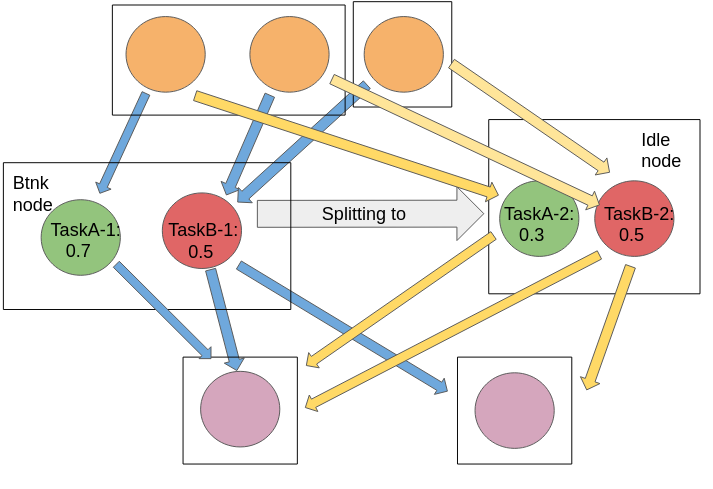}
    \caption{Example of SPLIT, the circles represent tasks, boxes represent nodes, yellow lines are newly incurred file transfers.}
\end{figure}

\begin{algorithm}
\caption{Node Splitting}
\begin{algorithmic}[1]
\renewcommand{\algorithmicrequire}{\textbf{Input:}}
\renewcommand{\algorithmicensure}{\textbf{Output:}}
\Require \\
Same as previous + an existing schedule
\Ensure map from nodes to scheduled tasks and portions  \\

\State $resTime\gets$ mapping from resources to their schedule times, initially given by input mapping
%\State $btnk \gets$ node to split
%\State $btnkTime \gets$ current bottleneck time
\While {$available Nodes \neq empty$}
\State $btnk \gets$ get bottleneck
\If{bottleneck is link}
\State $btnk\gets$  src or dst node, choose a batter one
\EndIf
\State $tasks \gets$ tasks on $btnk$ to be split
\State $pTasks$, $pNodes\gets$ parent tasks and nodes
\State $cTasks$, $cNodes\gets$ child tasks and nodes
\State {$tmp \gets \{\}$} \Comment{mapping from $proc$ to resulted new max schedule time over effected resources if $btnk$ is split on $proc$ 100\%}
\ForAll{$proc\in\mathit{availableNodes}$}
    \State $nTime \gets$ $\sum_{task \in tasks}exec[task][proc]+0$
    \State $lTimes \gets$ existing link usages in $resTime$
    \ForAll{$task \in tasks$}
        \ForAll{$pt \in pTasks$}
            \State $pNode\gets$ node of parent task $pt$
            \State $lk$ = $pNode$-$proc$ \Comment{parent link}
            \State $lTimes$[$lk$] += time of file $pt$-$task$ on $lk$ 
        \EndFor
        \ForAll{$ct \in cTasks$}
            \State do the same for child tasks
        \EndFor
    \EndFor
    \State $maxTime \gets$ max ($nTime$, max($lTimes.vals()$))
    \State $tmp$[$proc$] = $maxTime$
\EndFor
\State $P, T \gets$ (key, val) in $tmp$ with smallest value
\State split $btnk$ on $P$ with portion $ptn$ (TBD)
\State $ptn \gets (1-ptn) * btnkTime == ptn * T$
\State update resource usages
\EndWhile
\end{algorithmic}
\end{algorithm}

Lastly, there's a very important runtime problem the probability based load balancing algorithm didn't address. When splitting a task which has multiple parents (such as task3 in fig~1), all the parent output files associated with the same input instance have to be sent to the same replica of child, otherwise the child task cannot start. For example, for the earlier diamond DAG (fig~1) with 4 tasks, task3 has parents task1 and task2. If task3 has two replicas, task3-1, task3-2, then output files output-task1-input1 and output-task2-input1 must both be sent to either task3-1 or task3-2. Later, output-task1-input2 and output-task2-input2 have to follow the same rules, but they don't have to be sent to the same task3 replica with output-taskX-input1 files. For simplicity, in the real system, tasks are not aware of their peers (thus no distributed consensus needed). But if they only choose the child replica based on probability, then there's no guarantee that all parent output files associated with the same input instance will be sent to the same place. This only happens when a  multi-parent child task has more than one replica. The number of parents and the number of replicas for each parent doesn't matter, nor is this a problem if the child only has one parent task.

There are multiple solutions to this problem. For simplicity we chose the following. For single-parent tasks, we do the above probability based random selection; for multi-parent tasks, we extract the input task ID (tagged by the system), generate a hashcode using a self-defined hash function, then choose a child replica (rep) using 
\begin{equation}
    chosenRep = hashcode(input) \% numReps 
\end{equation}
An example hash function is simply atoi(input\_ID), which can result in a round-robin like distribution. Because same objects will always produce the same hashcodes, we can be sure any groups of intermediate files will not be scattered to different child replicas. As for hash collision, it doesn't hurt the correctness of the system, but we should try to distribute inputs as evenly as possible in order not to have a new bottleneck.

\subsection{Task Duplication (Algorithm Block 3)}
\label{sec: dup}

DUP is inspired by and adapted from theory-focused work~\cite{diyi}. It is designed to bypass slow links by rerouting the parent-child file transfer paths. As the name suggests, we're duplicating tasks, unlike in SPLIT, where we split nodes (all tasks on a node). We can see later that DUP would change the DAG itself as we modify the parent-child relations.

\begin{figure}[t]
    \centering
    \includegraphics[scale=0.32]{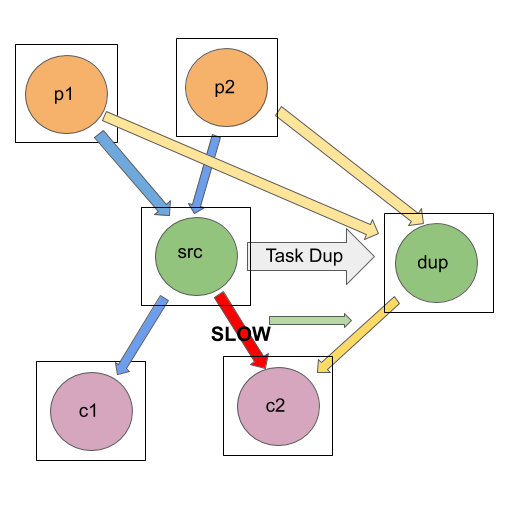}
    \caption{Example of duplication for a single task}
\end{figure}

The idea behind duplication comes from the scenario described in fig~4. Under a certain schedule, if a link is very slow due to network issues, we want to find a way to bypass the link by relocating the tasks whose file transfers go though the link. In fig~4, we want to bypass the slow red link. There's just one task that uses this link, namely src. In order to bypass the link, we need to relocate src. However, we shouldn't directly move it to a new node, because part of it doesn't go through the slow link (path \{p, src, c1\}). Our solution is to create a new task, which is same as src, namely dup, and place it on a selected node. With this, we redirected the traffic \{p, src, c2\} to \{p, dup, c2\}, but kept the traffic of \{p, src, c1\} that goes through the task but doesn't go through the bottleneck link.

Above is just a simple example of DUP where we used three parallel links to replace a slow link. In real-world cases, there are lots of other problems to consider.
\begin{itemize}
\item In DUP, using fig~4 as an example, we removed the parent-child relationships between src and c2, and added the relationship between p1 and dup, p2 and dup, dup and c2. Each round of duplication will change the DAG structure.
\item Unlike SPLIT, which in theory can always improve system performance if handled properly, duplication does not always improve performance. There are two conditions, first the bottleneck has to be a link, second the max schedule time on the new path has to be smaller than the bottleneck. 
\item TPHEFT and SPLIT are very general algorithms that can be applied anywhere, whereas DUP has several limitations. In DUP, we can only bypass slow links. But the getBottleneck() function doesn't know whether a link bottleneck is due to slow link or large file transfer. Applying DUP to the latter case can make things worse in runtime. 
\item Since DUP can change the parent-child relations, it can leave a trail of zombie tasks - the ones that are disconnected from the exit task, thus have no contribution to final output. We can set up a garbage collector to remove these tasks and their schedule to free up the nodes, and they should not be counted as bottlenecks.
\end{itemize}

This algorithm also consists of two phases, the find best node phase, which returns the current best available node; and the duplicate phase, which updates the nodes and links usages and the DAG. We use the same greedy method, at each iteration, choose an idle node which will have the best result and duplicate required tasks on it, until there are no resources, or duplication decreases throughput, or bottleneck is not a link anymore, as shown in the algorithm block. 

In the find best node phase, tasks to be duplicated are the ones on src node who transfer files to tasks on dst node. After tentatively choosing a candidate, the effected resources of duplication include, candidate node, link from candidate node to dst node, links from parent nodes of src node to candidate node. The resulted max schedule time across those resources have to be smaller than the bottleneck value.

\begin{algorithm}
\caption{Task Duplication}
\begin{algorithmic}[1]
\renewcommand{\algorithmicrequire}{\textbf{Input:}}
\renewcommand{\algorithmicensure}{\textbf{Output:}}
\Require \\
Same as previous + an existing schedule
\Ensure new mapping and DAG \\

\State $resTime\gets$ same meaning as previous
\While {$available Nodes \neq empty$ and $btnk$ is link}
\State $btnk \gets$ current bottleneck resource
\State $srcNode, dstNode\gets$ two sides of $btnk$
\State $srcTasks$, $dstTasks \gets$ tasks that use $btnk$
%\State $srcTasks$ are the set of tasks to be duplicated 
\State $pTasks, pNodes \gets$ parent tasks, nodes of $srcTasks$
\State $tmp \gets$ map from $proc$ to max schedule time over effected resources if duplicate $srcTasks$ on $proc$
\ForAll{$proc\in\mathit{availableNodes}$}
    \State $nTime \gets$ $\sum_{task \in srcTasks}exec[task][proc]+0$
    \State $lTimes \gets$ existing link usages in $resTime$
    \ForAll{$task \in srcTasks$}
        \ForAll{$pt \in pTasks$}
            \State $lk$ = $pNode$-$proc$ \Comment{parent link}
            \State $lTimes$[$lk$] += time of file $pt$-$task$ on $lk$ 
        \EndFor
        \ForAll{$ct \in cTasks$}
            \State $lk$ = $proc$-$dstNode$
            \State $lTimes$[$lk$] += time of file $task$-$ct$ on $lk$ 
        \EndFor
    \EndFor
    \State $maxTime \gets$ max ($nTime$, max($lTimes.vals()$))
    \State $tmp$[$proc$] = $maxTime$
\EndFor
\State $P, T \gets$ (key, val) in $tmp$ with smallest value
\State if $T$ \textgreater $btnkTime$ then  break
\State update usages (same as SPLIT) + reconstruct DAG

\EndWhile
\end{algorithmic}
\end{algorithm}

\section{System Implementation}

The original Jupiter is discussed in~\cite{ghosh2019jupiter}. In this paper, we built three new schedulers, execution engines (called CIRCE) and orchestration engines.

Jupiter (fig~5) is a containerized distributed system built on k8s, with network profilers, execution profilers, schedulers, execution and orchestration engines. Each profiler is a pod running on a k8s node, that collects computation and communication info and reports to the home node, where the scheduler pod is running. Then the scheduler calculates a mapping and sends it to CIRCE home pod via rest API. CIRCE home spawns a pod for each task (in SPLIT, a pod for each task replica), and ships the pod to the designated node. Pods can directly communicate with each other via clusterIP service. Each CIRCE pod has an input queue and output queue, whenever one or a group of (depending on task requirement) files is received in the input queue, it is used as input for that pod's task, and generates one or several files to the output queue. The output queue, upon receiving a file, decides which child pod to transfer the file to, either to a fixed child, or by random number generator or hashcode-bucket in case of SPLIT. Since we consider only single-core nodes, one node (not pod) can only be processing one instance at a time. At the node level, it looks like a FIFO queue.

\begin{figure}[htbp]
\includegraphics[scale=0.25]{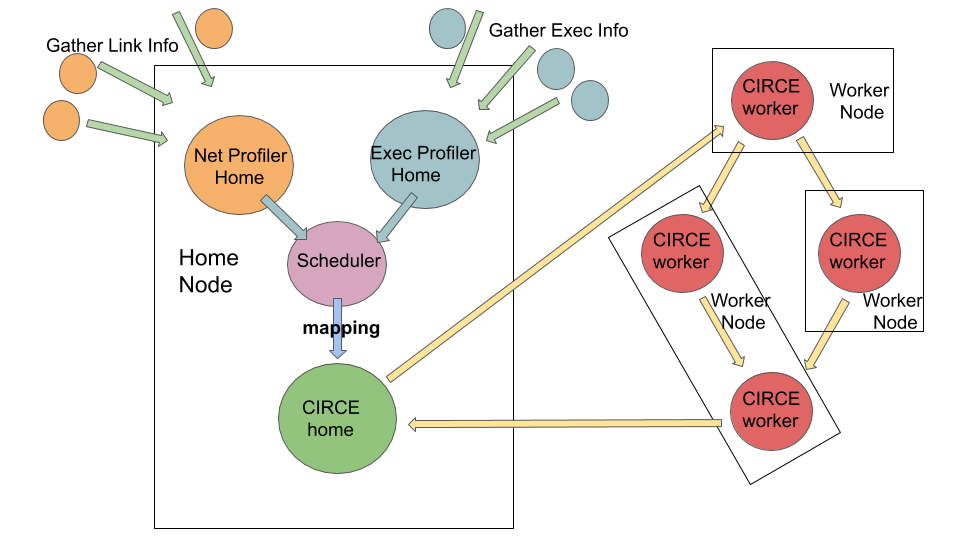}
\caption{Jupiter archetecture}
\end{figure}

In pipelined computing, a CIRCE pod can be receiving multiple files from multiple parents, and it has to wait for all parent output files of one particular instance to arrive before execution. But during this time period, it could receive parent files from different input instances. The system adds labels to intermediate files to indicate their associated input instances. When a file arrives, CIRCE will first cache its location and input instance. When the bucket of an input instance reaches designated size, those files are used as inputs for current task. 

A pod doesn't know the existence of parents or peers, it just monitors its input queues; but it knows everything about its children. An example of a pod's ENVs in SPLIT, which indicate two child pods, each with two replicas\\
$task0$-$1$/$0.15$ : $task0$-$2$/$0.85$ : $task1$-$1$/$0.5$ : $task1$-$2$/$0.5$ \\ 
In non hashnode-bucket case, this pod generates two random numbers under 1000. For the first number, if it's 0-149, transfer file to task0-1, else to task0-2. Similar for second number. In non-split cases, since there are no replicas, each replica has 100\% probability. Such a design greatly simplifies the system, with no consensus needed.

\section{Experiment Results}
\label{sec: exp}
\subsection{Experiment Setup}

We used a public cloud called Digital Ocean (DO) to conduct experiments. We designed 9 DAGs, 5 small dummy, 3 large dummy and 1 real application, with different shapes, sizes and parameters. For each DAG, we conduct 7 experiments, and measure the time of executing a few hundred instances in each experiment to calculate steady-state throughput. We first do HEFT, then TPHEFT to show improved throughput. Then we do HEFT+SPLIT and TPHEFT+SPLIT, to show that SPLIT works well on both baseline schedules, but has a better performance when used on an imbalanced schedule (HEFT). We also show that TPHEFT works best when the degree of parallelism inside the DAG (measure of how many tasks can be executed at the same time for the same input) is low.

To further show the impact of computation and communication parameters within the same DAG for SPLIT, we took the diamond shape DAG1, and did two sets of experiments. In the first set, we scaled the computation in each task by a factor of 1, 5, 10, 20, and measured the throughput improvement of HEFT+SPLIT over HEFT; in the second set, we scaled the communication in each transfer to a factor of 1, 5, 10, 20 and measured the improvement of MANUAL+SPLIT over MANUAL. Here the MANUAL is a schedule that places each task on a different node, thus forcibly creating several link bottlenecks to split.

We then show the improvement of DUP. This technique was invented to bypass slow links, so to show significant improvement, we simulate slow links by scheduling most tasks in US and one or a few in Europe (called MANUAL in table~III). The slow links will be bypassed by duplicating European tasks back to US. We established DO clusters that span across geographical data centers. Note that DUP can be applied to any schedule, but using a schedule where a slow link is a clear bottleneck is easy to show improvement. Additionally we also showed TPHEFT+DUP, but we don't expect it to give much improvement, because in an already balanced schedule such as TPHEFT, it's not so likely that links will be big bottlenecks. Our way of simulating slow links has a feature - when main data center is in the US, slow links always come in pairs because both source and destination are in US. This sometimes decreased performance. %, because for slow links, TPHEFT will bypass them; for large file transfers, TPHEFT will schedule parent and child on the same node, thus bypass the transfer. 

We considered but did not implement the combination of DUP and SPLIT, because: a) it is challenging from a system perspective to combine the two implementations and b) it is unclear that this combination would yield clear benefits. We leave open to future work to investigate this combination .

We use both dummy and real application DAGs for our experiments. The dummy application is called ``count", where each task does math calculations for a designated loop range. Such task mainly consumes CPU, and the execution time is easy to control by controlling the loop range. The real application we used is machine learning based object detection.

In below experiments, we show the number of input instances processed per 1000 seconds under steady-state. This is calculated by measuring the time of executing hundreds of instances. The percentage showed relative throughput increase before rounding, and the actual values are after rounding.

\subsection{Experiment Results}
\paragraph{Small dummy DAGs on small cluster, DAG1-5}
This part is done on DO, 1 master node, in SFO data center, with 4GB RAM and 2 cores; 15 worker nodes, with 1 GB RAM and 1 core, 8 in SFO, 3 in NYC, 3 in Germany. We used 5 random DAGs of different shapes and sizes (shown in fig~6). DAG1 and DAG2 are of same shape, but T1 in DAG2 is much heavier than T1 in DAG1 while other parameters are the same, thus creating balanced and imbalanced version of the same DAG.

\begin{figure}[t]
    \centering
    \includegraphics[scale=0.50]{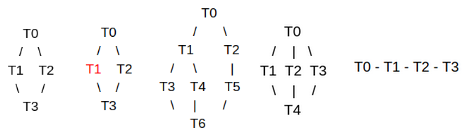}
    \caption{Small DAGs}
\end{figure}

\begin{table}[]
\caption{Throughput of HEFT, TPHEFT and their SPLIT}
\begin{tabular}{|l|l|l|l|l|}
\hline
 &
  \begin{tabular}[c]{@{}l@{}}HEFT\\ (vs None)\end{tabular} &
  \begin{tabular}[c]{@{}l@{}}HEFT+SPLIT\\ (vs HEFT)\end{tabular} &
  \begin{tabular}[c]{@{}l@{}}TPHEFT\\ (vs HEFT)\end{tabular} &
  \begin{tabular}[c]{@{}l@{}}TPHEFT+SPLIT\\ (vs TPHEFT)\end{tabular} \\ \hline
DAG1 &
  \begin{tabular}[c]{@{}l@{}}294\\ (+0.0\%)\end{tabular} &
  \begin{tabular}[c]{@{}l@{}}946\\ (+221.7\%)\end{tabular} &
  \begin{tabular}[c]{@{}l@{}}461\\ (+56.7\%)\end{tabular} &
  \begin{tabular}[c]{@{}l@{}}641\\ (+39.1\%)\end{tabular} \\ \hline
DAG2 &
  \begin{tabular}[c]{@{}l@{}}141\\ (+0.0\%)\end{tabular} &
  \begin{tabular}[c]{@{}l@{}}417\\ (+196.3\%)\end{tabular} &
  \begin{tabular}[c]{@{}l@{}}197\\ (+40.0\%)\end{tabular} &
  \begin{tabular}[c]{@{}l@{}}507\\ (+157.9\%)\end{tabular} \\ \hline
DAG3 &
  \begin{tabular}[c]{@{}l@{}}392\\ (+0.0\%)\end{tabular} &
  \begin{tabular}[c]{@{}l@{}}521\\ (+32.8\%)\end{tabular} &
  \begin{tabular}[c]{@{}l@{}}472\\ (+20.3\%)\end{tabular} &
  \begin{tabular}[c]{@{}l@{}}552\\ (+17.1\%)\end{tabular} \\ \hline
DAG4 &
  \begin{tabular}[c]{@{}l@{}}794\\ (+0.0\%)\end{tabular} &
  \begin{tabular}[c]{@{}l@{}}1458\\ (+83.7\%)\end{tabular} &
  \begin{tabular}[c]{@{}l@{}}1282\\ (+61.5\%)\end{tabular} &
  \begin{tabular}[c]{@{}l@{}}1961\\ (+52.9\%)\end{tabular} \\ \hline
DAG5 &
  \begin{tabular}[c]{@{}l@{}}521\\ (+0.0\%)\end{tabular} &
  \begin{tabular}[c]{@{}l@{}}1736\\ (+233.3\%)\end{tabular} &
  \begin{tabular}[c]{@{}l@{}}1299\\ (+149.4\%)\end{tabular} &
  \begin{tabular}[c]{@{}l@{}}2326\\ (+79.1\%)\end{tabular} \\ \hline
DAG6 &
  \begin{tabular}[c]{@{}l@{}}315\\ (+0.0\%)\end{tabular} &
  \begin{tabular}[c]{@{}l@{}}459\\ (+45.5\%)\end{tabular} &
  \begin{tabular}[c]{@{}l@{}}671\\ (+112.8\%)\end{tabular} &
  \begin{tabular}[c]{@{}l@{}}794\\ (+18.3\%)\end{tabular} \\ \hline
DAG7 &
  \begin{tabular}[c]{@{}l@{}}228\\ (+0.0\%)\end{tabular} &
  \begin{tabular}[c]{@{}l@{}}272\\ (+19.2\%)\end{tabular} &
  \begin{tabular}[c]{@{}l@{}}758\\ (+231.8\%)\end{tabular} &
  \begin{tabular}[c]{@{}l@{}}935\\ (+23.4\%)\end{tabular} \\ \hline
DAG8 &
  \begin{tabular}[c]{@{}l@{}}284\\ (+0.0\%)\end{tabular} &
  \begin{tabular}[c]{@{}l@{}}427\\ (+50.4\%)\end{tabular} &
  \begin{tabular}[c]{@{}l@{}}625\\ (+120.0\%)\end{tabular} &
  \begin{tabular}[c]{@{}l@{}}510\\ (-18.4\%)\end{tabular} \\ \hline
IMGP &
  \begin{tabular}[c]{@{}l@{}}498\\ (+0.0\%)\end{tabular} &
  \begin{tabular}[c]{@{}l@{}}763\\ (+53.4\%)\end{tabular} &
  \begin{tabular}[c]{@{}l@{}}524\\ (+5.0\%)\end{tabular} &
  \begin{tabular}[c]{@{}l@{}}1000\\ (+91.0\%)\end{tabular} \\ \hline
\end{tabular}
\end{table}

Note that this part only refers to DAG1 to DAG5. It is not meaningful to compare results across DAG1-5 and DAG6-8, because they're done on different clusters. In table~II, we see that TPHEFT (compared with HEFT) works best for DAGs with low degrees of parallelism. Linear DAGs (DAG5) have no parallelism, thus most improvement. DAG3 has high parallelism compared with the other four, thus a low improvement. This is expected from our previous analysis.

Also, SPLIT works best for imbalanced schedules. We see from table~II that when used upon HEFT, SPLIT gives higher throughput improvement than when used upon TPHEFT, because the latter is a more balanced schedule. Another case we mentioned earlier is when one or a few tasks are significantly heavier than others. Compare DAG1 and DAG2, we see that for TPHEFT baseline, DAG2's SPLIT showed more improvement because it's imbalanced compared with DAG1. It's not the same case for HEFT+SPLIT between DAG1 and DAG2, i.e. though DAG2 is more imbalanced, the SPLIT on HEFT and DAG2 was not so significant compared with SPLIT on HEFT and DAG1. This is because the imbalance of task T1 is outweighed by the imbalance of HEFT.

For scaling computation and communication, results are shown in fig~11 and fig~12. As we see, with bigger scale factor, SPLIT gives better performance. This is due to higher degree of imbalance. For communication, SPLIT doesn't perform as well as for computation, because of the limitations of our model, but it still gives increasing performance.

Looking at DUP (table~III), we see for most DAGs it resulted in good improvement compared with the manual schedule where cross continental links are big bottlenecks, except in DAG2, where task1 is significantly heavier than other tasks, and the bottleneck is not on links, in which case DUP can't improve overall system performance. This further shows where DUP really fits. Also, we see for TPHEFT, DUP has no improvement. Because we used public clouds, the real links are always good, besides TPHEFT will always try to avoid link bottltnecks anyway, thus it's hard to create or simulate a case where TPHEFT+DUP is any good.

As we mentioned earlier, there are multiple factors that might make our system model different from the real world case, including virtual link model; multi-socket communication cost; etc. Although we don't know exactly how much of an impact each one has, we can compare the compile time throughput with runtime ones. Our results of these particular experiment show the value ranges from 1.4 to 3.5. A major reason is the virtual link model. Though our model values are different from real values, we actually improved real values significantly by improving model values.

\paragraph{Large DAGs on large cluster, DAG6-9}
In this experiment, we use a 40 node DO cluster, with 30 nodes in SFO and 10 nodes in Europe. We chose some other DAGs illustrated in fig~7 (DAG6), fig~9 (DAG7), fig~10 (DAG8), and do similar experiments and summarize them in tables. These DAGs have moderate sizes and application backgrounds, although we're still running the dummy app.

\begin{figure}[htbp]
\begin{minipage}[t]{0.5\linewidth}
    \includegraphics[width=\linewidth]{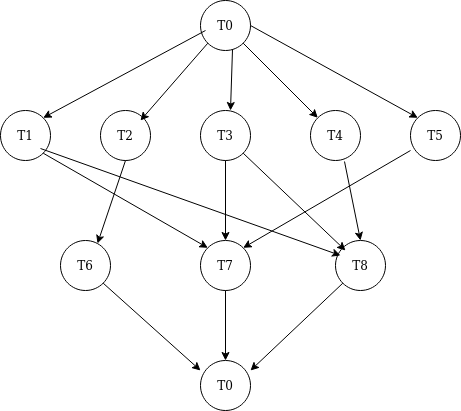}
    \caption{DAG6}
    \label{f1}
\end{minipage}%
    \hfill%
\begin{minipage}[t]{0.45\linewidth}
    \includegraphics[width=\linewidth]{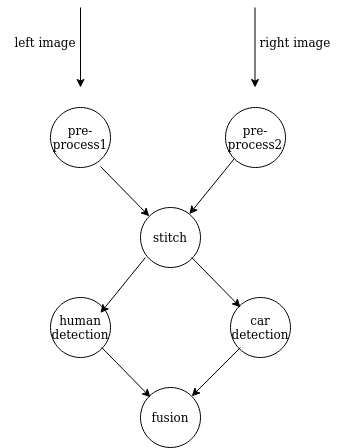}
    \caption{Image Processing}
    \label{f2}
\end{minipage} 
\end{figure}

\begin{figure}[htbp]
\begin{minipage}[t]{0.55\linewidth}
    \includegraphics[width=\linewidth]{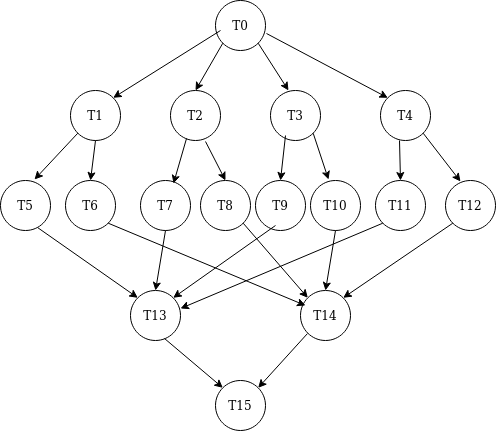}
    \caption{DAG7}
    \label{f1}
\end{minipage}%
    \hfill%
\begin{minipage}[t]{0.28\linewidth}
    \includegraphics[width=\linewidth]{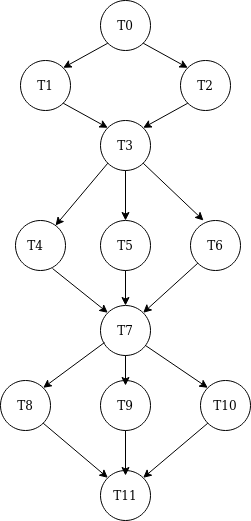}
    \caption{DAG8}
    \label{f2}
\end{minipage} 
\end{figure}

\begin{figure}[htbp]
\begin{minipage}[t]{0.5\linewidth}
    \includegraphics[width=\linewidth]{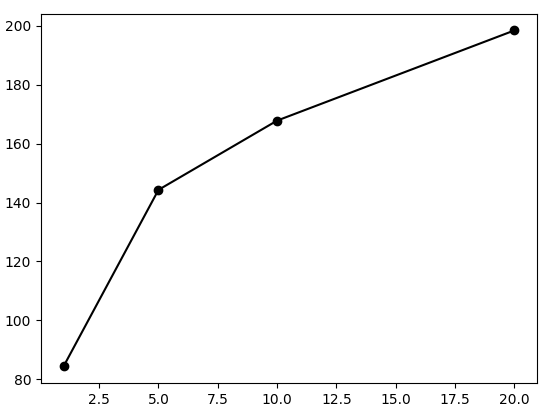}
    \caption{TP increase (\%) vs\\ computation scale factor}
    \label{f1}
\end{minipage}%
    \hfill%
\begin{minipage}[t]{0.5\linewidth}
    \includegraphics[width=\linewidth]{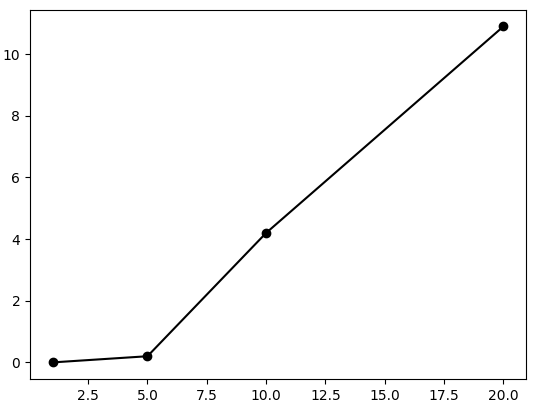}
    \caption{TP increase (\%) vs\\ communication scale factor}
    \label{f2}
\end{minipage} 
\end{figure}

It is worth noting that in DAG8, although TPHEFT improved a lot on HEFT, SPLIT decreased throughput when applied on TPHEFT. We analyzed the data. Our conclusion is, for SPLIT, when computation and communication costs are close, the actual system bottleneck is a link, but since we used virtual links, one actual link may be split into several virtual links, thus none of them was counted as bottleneck. Instead we end up splitting nodes, which are not actual bottlenecks. With additional overhead, the more we split, the lower the throughput. For link bottlenecks, different clusters require different models, unless we know the link topology, we don't know what's the best model for a particular cluster. But as a general algorithm, we can keep the virtual link model and consider rounds of splitting as a hyper-parameter and try different values including zero, to get the best results. But this only happens because TPHEFT results in a well-balanced allocation. SPLIT still works well on HEFT.

\begin{table}[]
\caption{Throughput of MANUAL, TPHEFT and DUP}
\begin{tabular}{|l|l|l|l|l|}
\hline
 &
  \begin{tabular}[c]{@{}l@{}}MANUAL\\ (vs None)\end{tabular} &
  \begin{tabular}[c]{@{}l@{}}MANUAL+DUP\\ (vs MANUAL)\end{tabular} &
  \begin{tabular}[c]{@{}l@{}}TPHEFT\\ (vs None)\end{tabular} &
  \begin{tabular}[c]{@{}l@{}}TPHEFT+DUP\\ (vs TPHEFT)\end{tabular} \\ \hline
DAG1 &
  \begin{tabular}[c]{@{}l@{}}312\\ (+0.0\%)\end{tabular} &
  \begin{tabular}[c]{@{}l@{}}549\\ (+76.0\%)\end{tabular} &
  \begin{tabular}[c]{@{}l@{}}461\\ (+56.7\%)\end{tabular} &
  \begin{tabular}[c]{@{}l@{}}461\\ (+0.0\%)\end{tabular} \\ \hline
DAG2 &
  \begin{tabular}[c]{@{}l@{}}172\\ (+0.0\%)\end{tabular} &
  \begin{tabular}[c]{@{}l@{}}169\\ (-1.7\%)\end{tabular} &
  \begin{tabular}[c]{@{}l@{}}197\\ (+40.0\%)\end{tabular} &
  \begin{tabular}[c]{@{}l@{}}197\\ (+0.0\%)\end{tabular} \\ \hline
DAG3 &
  \begin{tabular}[c]{@{}l@{}}287\\ (+0.0\%)\end{tabular} &
  \begin{tabular}[c]{@{}l@{}}413\\ (+43.9\%)\end{tabular} &
  \begin{tabular}[c]{@{}l@{}}472\\ (+20.3\%)\end{tabular} &
  \begin{tabular}[c]{@{}l@{}}472\\ (+0.0\%)\end{tabular} \\ \hline
DAG4 &
  \begin{tabular}[c]{@{}l@{}}833\\ (+0.0\%)\end{tabular} &
  \begin{tabular}[c]{@{}l@{}}1064\\ (+27.7\%)\end{tabular} &
  \begin{tabular}[c]{@{}l@{}}1282\\ (+61.5\%)\end{tabular} &
  \begin{tabular}[c]{@{}l@{}}1210\\ (-5.6\%)\end{tabular} \\ \hline
DAG5 &
  \begin{tabular}[c]{@{}l@{}}694\\ (+0.0\%)\end{tabular} &
  \begin{tabular}[c]{@{}l@{}}1000\\ (+54.1\%)\end{tabular} &
  \begin{tabular}[c]{@{}l@{}}1299\\ (+149.4\%)\end{tabular} &
  \begin{tabular}[c]{@{}l@{}}1197\\ (-7.9\%)\end{tabular} \\ \hline
DAG6 &
  \begin{tabular}[c]{@{}l@{}}476\\ (+0.0\%)\end{tabular} &
  \begin{tabular}[c]{@{}l@{}}532\\ (+11.8\%)\end{tabular} &
  \begin{tabular}[c]{@{}l@{}}671\\ (+112.8\%)\end{tabular} &
  \begin{tabular}[c]{@{}l@{}}671\\ (+0.0\%)\end{tabular} \\ \hline
DAG7 &
  \begin{tabular}[c]{@{}l@{}}266\\ (+0.0\%)\end{tabular} &
  \begin{tabular}[c]{@{}l@{}}304\\ (+14.3\%)\end{tabular} &
  \begin{tabular}[c]{@{}l@{}}758\\ (+231.8\%)\end{tabular} &
  \begin{tabular}[c]{@{}l@{}}758\\ (+0.0\%)\end{tabular} \\ \hline
DAG8 &
  \begin{tabular}[c]{@{}l@{}}316\\ (+0.0\%)\end{tabular} &
  \begin{tabular}[c]{@{}l@{}}302\\ (-4.4\%)\end{tabular} &
  \begin{tabular}[c]{@{}l@{}}625\\ (+120.0\%)\end{tabular} &
  \begin{tabular}[c]{@{}l@{}}625\\ (+0.0\%)\end{tabular} \\ \hline
IMGP &
  \begin{tabular}[c]{@{}l@{}}500\\ (+0.0\%)\end{tabular} &
  \begin{tabular}[c]{@{}l@{}}556\\ (+11.2\%)\end{tabular} &
  \begin{tabular}[c]{@{}l@{}}524\\ (+5.0\%)\end{tabular} &
  \begin{tabular}[c]{@{}l@{}}524\\ (+0.0\%)\end{tabular} \\ \hline
\end{tabular}
\end{table}

\paragraph{Machine learning based object detection application}
To show an example of real-world application with our system, we composed an object detection application (fig~8), which is a good example of dispersed computing. The input images are taken by different cameras from different angles, and we want detect objects in the combined image, which gives an overview of the scene. The app tasks two images (left and right image, with overlapping parts) as inputs, pre-processes each image (such as resize), then sends these two processed images to the stitching task, where we find the overlapping part to stitch the two processed images together. After this, we use a pre-trained model to locate humans and cars in the stitched image. In the end we group all the results as one matrix in a csv file as output. Throughput results are shown in the two tables in the ``IMGP" row. They're basically consistent with results in the dummy DAG, given the specific shape and parameters of the IMGP application.

\section{Conclusion and Future Work}
\label{sec: dis-distributed}

In this paper, we proposed three throughput optimized scheduling algorithms for dispersed computing, and implemented a real-world distributed system to support them. We further conducted experiments using different DAGs, both dummy and real, to show the throughput improvement under different circumstances. TPHEFT and SPLIT are common algorithms while DUP is specially designed for slow links. 
In the future, we can extend our algorithms to non-idle nodes, multi-core CPUs, different models for link topology, and different applications from compute oriented ones. Further, we plan to work on dynamic scheduling with machine learning based approaches.

\section*{Acknowledgment}

This material is based upon work supported by Defense Advanced Research Projects Agency (DARPA) under Contract No. HR001117C0053. Any views, opinions, and/or findings expressed are those of the author(s) and should not be interpreted as representing the official views or policies of the Department of Defense or the U.S. Government.



\bibliographystyle{IEEEtran}
\bibliography{IEEEabrv,citation_total}

\begin{thebibliography}{10}
\providecommand{\url}[1]{#1}
\csname url@samestyle\endcsname
\providecommand{\newblock}{\relax}
\providecommand{\bibinfo}[2]{#2}
\providecommand{\BIBentrySTDinterwordspacing}{\spaceskip=0pt\relax}
\providecommand{\BIBentryALTinterwordstretchfactor}{4}
\providecommand{\BIBentryALTinterwordspacing}{\spaceskip=\fontdimen2\font plus
\BIBentryALTinterwordstretchfactor\fontdimen3\font minus
  \fontdimen4\font\relax}
\providecommand{\BIBforeignlanguage}[2]{{%
\expandafter\ifx\csname l@#1\endcsname\relax
\typeout{** WARNING: IEEEtran.bst: No hyphenation pattern has been}%
\typeout{** loaded for the language `#1'. Using the pattern for}%
\typeout{** the default language instead.}%
\else
\language=\csname l@#1\endcsname
\fi
#2}}
\providecommand{\BIBdecl}{\relax}
\BIBdecl

\bibitem{19middleware}
\BIBentryALTinterwordspacing
P.~Ghosh, Q.~Nguyen, and B.~Krishnamachari, ``Container orchestration for
  dispersed computing,'' in \emph{Proceedings of the 5th International Workshop
  on Container Technologies and Container Clouds}, ser. WOC '19.\hskip 1em plus
  0.5em minus 0.4em\relax New York, NY, USA: Association for Computing
  Machinery, 2019, p. 19–24. [Online]. Available:
  \url{https://doi.org/10.1145/3366615.3368354}
\BIBentrySTDinterwordspacing

\bibitem{dcomp_intro}
M.~R. {Schurgot}, M.~{Wang}, A.~E. {Conway}, L.~G. {Greenwald}, and P.~D.
  {Lebling}, ``A dispersed computing architecture for resource-centric
  computation and communication,'' \emph{IEEE Communications Magazine},
  vol.~57, no.~7, pp. 13--19, 2019.

\bibitem{coded_dispersed}
\BIBentryALTinterwordspacing
S.~Li, M.~A. Maddah{-}Ali, and A.~S. Avestimehr, ``Coding for distributed fog
  computing,'' \emph{CoRR}, vol. abs/1702.06082, 2017. [Online]. Available:
  \url{http://arxiv.org/abs/1702.06082}
\BIBentrySTDinterwordspacing

\bibitem{social_dispersed}
M.~Garc{\'\i}a-Valls, A.~Dubey, and V.~Botti, ``Introducing the new paradigm of
  social dispersed computing: applications, technologies and challenges,''
  \emph{Journal of Systems Architecture}, vol.~91, pp. 83--102, 2018.

\bibitem{scheduling_cloud}
L.~F. Bittencourt, A.~Goldman, E.~R. Madeira, N.~L. da~Fonseca, and
  R.~Sakellariou, ``Scheduling in distributed systems: A cloud computing
  perspective,'' \emph{Computer science review}, vol.~30, pp. 31--54, 2018.

\bibitem{fault_tolerance}
F.~C. G{\"a}rtner, ``Fundamentals of fault-tolerant distributed computing in
  asynchronous environments,'' \emph{ACM Computing Surveys (CSUR)}, vol.~31,
  no.~1, pp. 1--26, 1999.

\bibitem{mobile_benchmark}
A.~Ignatov, R.~Timofte, W.~Chou, K.~Wang, M.~Wu, T.~Hartley, and L.~Van~Gool,
  ``Ai benchmark: Running deep neural networks on android smartphones,'' in
  \emph{Proceedings of the European conference on computer vision (ECCV)},
  2018, pp. 0--0.

\bibitem{cpu_benchmark}
C.~Byun, J.~Kepner, W.~Arcand, D.~Bestor, B.~Bergeron, V.~Gadepally, M.~Houle,
  M.~Hubbell, M.~Jones, A.~Klein \emph{et~al.}, ``Benchmarking data analysis
  and machine learning applications on the intel knl many-core processor,''
  \emph{arXiv preprint arXiv:1707.03515}, 2017.

\bibitem{Jupiter_Quynh}
Q.~Nguyen, P.~Ghosh, and B.~Krishnamachari, ``End-to-end network performance
  monitoring for dispersed computing,'' March 2018.

\bibitem{heft}
H.~{Topcuoglu}, S.~{Hariri}, and {Min-You Wu}, ``Performance-effective and
  low-complexity task scheduling for heterogeneous computing,'' \emph{IEEE
  Transactions on Parallel and Distributed Systems}, vol.~13, no.~3, pp.
  260--274, 2002.

\bibitem{borg}
\BIBentryALTinterwordspacing
A.~Verma, L.~Pedrosa, M.~Korupolu, D.~Oppenheimer, E.~Tune, and J.~Wilkes,
  ``Large-scale cluster management at google with borg,'' in \emph{Proceedings
  of the Tenth European Conference on Computer Systems}, ser. EuroSys
  '15.\hskip 1em plus 0.5em minus 0.4em\relax New York, NY, USA: Association
  for Computing Machinery, 2015. [Online]. Available:
  \url{https://doi.org/10.1145/2741948.2741964}
\BIBentrySTDinterwordspacing

\bibitem{npcomplete}
J.~D. Ullman, ``Np-complete scheduling problems,'' \emph{Journal of Computer
  and System sciences}, vol.~10, no.~3, pp. 384--393, 1975.

\bibitem{cluster_tpds}
H.~{Kanemitsu}, M.~{Hanada}, and H.~{Nakazato}, ``Clustering-based task
  scheduling in a large number of heterogeneous processors,'' \emph{IEEE
  Transactions on Parallel and Distributed Systems}, vol.~27, no.~11, pp.
  3144--3157, 2016.

\bibitem{cluster2}
P.~{Neamatollahi}, S.~{Abrishami}, M.~{Naghibzadeh}, M.~H. {Yaghmaee
  Moghaddam}, and O.~{Younis}, ``Hierarchical clustering-task scheduling policy
  in cluster-based wireless sensor networks,'' \emph{IEEE Transactions on
  Industrial Informatics}, vol.~14, no.~5, pp. 1876--1886, 2018.

\bibitem{diyi}
D.~{Hu} and B.~{Krishnamachari}, ``Throughput optimized scheduler for dispersed
  computing systems,'' in \emph{2019 7th IEEE International Conference on
  Mobile Cloud Computing, Services, and Engineering (MobileCloud)}, 2019, pp.
  76--84.

\bibitem{cluster_icpp}
D.~Bozdag, F.~Ozguner, E.~Ekici, and U.~Catalyurek, ``A task duplication based
  scheduling algorithm using partial schedules,'' in \emph{2005 International
  Conference on Parallel Processing (ICPP'05)}.\hskip 1em plus 0.5em minus
  0.4em\relax IEEE, 2005, pp. 630--637.

\bibitem{ipdps09}
M.~Gallet, L.~Marchal, and F.~Vivien, ``Efficient scheduling of task graph
  collections on heterogeneous resources,'' in \emph{2009 IEEE International
  Symposium on Parallel \& Distributed Processing}.\hskip 1em plus 0.5em minus
  0.4em\relax IEEE, 2009, pp. 1--11.

\bibitem{odessa}
\BIBentryALTinterwordspacing
M.-R. Ra, A.~Sheth, L.~Mummert, P.~Pillai, D.~Wetherall, and R.~Govindan,
  ``Odessa: Enabling interactive perception applications on mobile devices,''
  in \emph{Proceedings of the 9th International Conference on Mobile Systems,
  Applications, and Services}, ser. MobiSys '11.\hskip 1em plus 0.5em minus
  0.4em\relax New York, NY, USA: Association for Computing Machinery, 2011, p.
  43–56. [Online]. Available: \url{https://doi.org/10.1145/1999995.2000000}
\BIBentrySTDinterwordspacing

\bibitem{linearDAG}
C.~{Yang}, R.~{Pedarsani}, and A.~S. {Avestimehr}, ``Communication-aware
  scheduling of serial tasks for dispersed computing,'' \emph{IEEE/ACM
  Transactions on Networking}, vol.~27, no.~4, pp. 1330--1343, 2019.

\bibitem{mapreduce}
J.~Dean and S.~Ghemawat, ``Mapreduce: Simplified data processing on large
  clusters,'' in \emph{OSDI'04: Sixth Symposium on Operating System Design and
  Implementation}, San Francisco, CA, 2004, pp. 137--150.

\bibitem{theory}
\BIBentryALTinterwordspacing
K.~Agrawal, J.~Li, K.~Lu, and B.~Moseley, ``Brief announcement: Scheduling
  parallelizable jobs online to maximize throughput,'' in \emph{Proceedings of
  the 29th ACM Symposium on Parallelism in Algorithms and Architectures}, ser.
  SPAA '17.\hskip 1em plus 0.5em minus 0.4em\relax New York, NY, USA:
  Association for Computing Machinery, 2017, p. 87–89. [Online]. Available:
  \url{https://doi.org/10.1145/3087556.3087590}
\BIBentrySTDinterwordspacing

\bibitem{simu}
Y.~{Gu} and Q.~{Wu}, ``Maximizing workflow throughput for streaming
  applications in distributed environments,'' in \emph{2010 Proceedings of 19th
  International Conference on Computer Communications and Networks}, 2010, pp.
  1--6.

\bibitem{dup_tp_stream}
\BIBentryALTinterwordspacing
N.~Vydyanathan, U.~Catalyurek, T.~Kurc, P.~Sadayappan, and J.~Saltz, ``A
  duplication based algorithm for optimizing latency under throughput
  constraints for streaming workflows,'' in \emph{Proceedings of the 2008 37th
  International Conference on Parallel Processing}, ser. ICPP '08.\hskip 1em
  plus 0.5em minus 0.4em\relax USA: IEEE Computer Society, 2008, p. 254–261.
  [Online]. Available: \url{https://doi.org/10.1109/ICPP.2008.68}
\BIBentrySTDinterwordspacing

\bibitem{lat_tp}
\BIBentryALTinterwordspacing
------, ``Optimizing latency and throughput of application workflows on
  clusters,'' \emph{Parallel Comput.}, vol.~37, no. 10–11, p. 694–712, Oct.
  2011. [Online]. Available: \url{https://doi.org/10.1016/j.parco.2010.05.003}
\BIBentrySTDinterwordspacing

\bibitem{ghosh2019jupiter}
P.~Ghosh, Q.~Nguyen, P.~K. Sakulkar, A.~Knezevic, J.~A. Tran, J.~Wang, Z.~Lin,
  B.~Krishnamachari, M.~Annavaram, and S.~Avestimehr, ``Jupiter: A networked
  computing architecture,'' 2019.

\end{thebibliography}
%\begin{thebibliography}{1}


%\end{thebibliography}

\end{document}